# MCSCF-DFT based on an interacting reference system.

Shusuke Yamanaka[1], Koichi Kusakabe[2], Kazuto Nakata[1], Toshikazu Takada[3],
Kizashi Yamaguchi[1]

1 Department of Chemistry, Faculty of Science, Osaka University, Machikaneyama 1-1, Toyonaka 560-0043, Japan

2 Department of Chemistry, Graduate School of Engineering Science, Osaka University, Machikaneyama 1-3, Toyonaka 560-8531, Japan

3 NEC Corporation, Miyukigaoka, Tsukuba, Ibaraki 305-8501, Japan

**Abstract**

We present the MCSCF version of density functional theory. Two sets of equations, which correspond to the CI and orbital relaxation respectively, are derived. An important feature is that the correlation potential of DFT for CI wavefunction and that for orbitals are different to each other. These relations ensure that a density yielded by an effective MCSCF solution also satisfies the Euler equation of DFT.

## Introduction

The Kohn-Sham (KS) density functional theory (DFT) [1,2] is now a powerful tool in computational chemistry. But it is known that it does not work well for near degeneracy of several configurations. In particular, the density that is not noninteracting v-representable is beyond the scope of KS-DFT [3]. Since this defect is due to not only the approximation of exchange-correlation functional but also the single determinant feature of KS-DFT, one possible remedy is to revive a multireference (MR) wavefucntion (WF) and to combine it with DFT. The first generation to merge MR-WF and DFT remains a double counting problem of electron correlation [4]. Although the prescription to cover the electron correlation by KS-DFT is quite different to that of the MR-WF theory, recent developers resolve this problem by various ways [5,6]. The definition of the spin-polarizability for a singlet MR-WF is also refined from the computational viewpoint [7].

As another direction, recent developments of MR-DFT are obviously towards the multireference version of density functional theory. Kusakabe introduced MR-WF in order to define the universal functional of the partially interacting system [8], deriving the set of equations called "the extended KS-DFT (EKS-DFT)". Savin and his co-workers have developed such a MR-DFT based on adiabatic connection by dividing the Coulomb interaction term in Hamiltonian into the short-range and long-range interaction [6]. We have recently developed an iterative CASCI-DFT (ICASCI-DFT) method [9]. The mathematical framework of ICASCI-DFT is given by a CASCI equation with a residual correlation potential, so its variational space of ICASCI-DFT is same to that of the CASCI approach. In addition, from the computational point of view, the orbital transformation procedure is skipped except that in the first cycle. Nevertheless, we can handle both nondynamical and dynamical correlation effects.

However, when an appropriate set of molecular orbitals is not available, the orbital relaxation procedure becomes essential to obtain the good density. Thus, in this study, we present the general formulation of the MCSCF-DFT. Roos and his co-workers have developed a precursive approach in this line [10]. They intend to add a total correction of DFT to CASSCF wavefunction approach that is pioneered by them, in order to obtain the any physical properties. This is a straightforward and probably powerful way to realize the CASSCF-DFT approach. The different point of our formalism compared with them is that the emphasis of our formalism is on the multireference implementation of DFT, in other words, the equivalent formalism of the Euler equation, which is the original equation of DFT. This spirit is of EKS theory, and leads us to our original relation between correlation functional and its potential given below.

**Effective CI equation for MR-DFT**

We start from the division of the energy into three parts,

$$E[\rho(\mathbf{r})] = Min_{\rho(\mathbf{r}) \to N} \left[ F^p[\rho(\mathbf{r})] + E_{RC}[\rho(\mathbf{r})] + \int d\mathbf{r} \rho(\mathbf{r}) V_{ext}(\mathbf{r}) \right], \quad (1a)$$

where first, second, and third terms in parenthesis at the right side are the modified universal functional residual correlation, and external potential terms, respectively. The modified universal functional is defined for the variational space of MCSCF wavefunctions:

$$F^p[\rho(\mathbf{r})] = Min^p_{\Psi \to \rho(\mathbf{r})} \langle \Psi | \hat{T} + \hat{V}_{ee} | \Psi \rangle. \quad (1b)$$

The variation of CI coefficients lead to the equation

$$\delta E[\rho(\mathbf{r})] = \sum_{ij} H^{core}_{ji} \delta P_{ij} + \sum_{ijkl} \langle kl|ij \rangle \delta \Pi_{ijkl} + \int d\mathbf{r} \frac{\delta E_{RC}}{\delta \rho}(\mathbf{r}) \delta \rho(\mathbf{r})$$

$$= \sum_{ij} H^{core}_{ji} \delta P_{ij} + \sum_{ijkl} \langle kl|ij \rangle \delta \Pi_{ijkl} + \sum_{ij} \langle i | \frac{\delta E_{RC}}{\delta \rho} | j \rangle \delta P_{ji} \quad (1c)$$

where $H^{core}_{ji}$, $P_{ji}$, $\Pi_{ijkl}$ are the core Hamiltonian, a spinless-one particle density matrix (1-DM), a spinless two-particle density matrix (2-DM), respectively, and the physicists' notation is used for electron-repulsion integrals. Here it is arbitrary whether the two-electron integrals are of atomic orbitals or of molecular orbitals. We here assume the latter, according to the usual convention of the MR-WF formalism [11]. It should be noted that the deviation given in eq.(1c) is limited within the specific MCSCF variational space. We now intend to formulate the effective multireference equation in the form of

$$\left(\hat{H}^{core} + \hat{V}_1^{eff} + \hat{V}_2^{eff}\right)|\Psi\rangle = E^{eff}|\Psi\rangle. \quad (2a)$$

The effective energy is thus given by,

$$E^{eff} = \langle\Psi|\left(\hat{H}^{core} + \hat{V}_1^{eff} + \hat{V}_2^{eff}\right)|\Psi\rangle$$
$$= \sum_{ij}\left(H_{ij}^{core} + V_{1,ij}^{eff}\right)P_{ij} + \sum_{ijkl}V_{2,ijkl}^{eff}\Pi_{ijkl} \quad (2b)$$

The deviation of the effective energy due to the variation of CI coefficients is given by

$$\delta E^{eff} = \sum_{ij}\left\{\left(H_{ij}^{core} + V_{1\,ij}^{eff}\right)\delta P_{ij} + \delta V_{1\,ij}^{eff} P_{ij}\right\} + \sum_{ijkl}\left(V_{2\,ijkl}^{eff}\delta\Pi_{ijkl} + \delta V_{2\,ijkl}^{eff}\Pi_{ijkl}\right). \quad (2c)$$

The effective CI wavefucntion for the ground state is obtained by the relation,

$$\sum_{ij}\left\{\left(H_{ij}^{core} + V_{1\,ij}^{eff}\right)\delta P_{ij} + \delta V_{1\,ij}^{eff} P_{ij}\right\} + \sum_{ijkl}\left(V_{2\,ijkl}^{eff}\delta\Pi_{ijkl} + \delta V_{2\,ijkl}^{eff}\Pi_{ijkl}\right) = 0. \quad (2d)$$

Comparing eq. (2b) with (1c), we found that if we set

$$V_{2\,ijkl}^{eff} = \langle kl|ij\rangle \quad (3a)$$

$$V_{1\,ij}^{eff} = \langle i|\hat{V}^{RC}|j\rangle \quad (3b)$$

with

$$E_{RC}[\rho(\mathbf{r})] = \int d\mathbf{r}V^{RC}(\mathbf{r})\rho(\mathbf{r}), \quad (3c)$$

the Euler equation is satisfied by the solution of effective MR-DFT given by eq. (2a).

To prove this relation, all we need is to substitute eqs. (3) into eq. (2c):

$$\delta E^{eff} = \sum_{ij}\left\{\left(H_{ij}^{core} + V_{1\,ij}^{eff}\right)\delta P_{ij} + \delta V_{1\,ij}^{eff} P_{ij}\right\} + \sum_{ijkl}\langle kl|ij\rangle\delta\Pi_{ijkl}$$
$$= \sum_{ij}H_{ij}^{core}\delta P_{ij} + \sum_{ijkl}\langle kl|ij\rangle\delta\Pi_{ijkl} + \int d\mathbf{r}\left[V^{RC}(\mathbf{r})\delta\rho(\mathbf{r}) + \delta V^{RC}(\mathbf{r})\rho(\mathbf{r})\right].$$
$$= \sum_{ij}H_{ij}^{core}\delta P_{ij} + \sum_{ijkl}\langle kl|ij\rangle\delta\Pi_{ijkl} + \int d\mathbf{r}\frac{\delta E^{RC}(\mathbf{r})}{\delta\rho(\mathbf{r})}\delta\rho(\mathbf{r})$$

Here, we use the fact that the deviation of density is given by

$$\delta\rho(\mathbf{r}) = \sum_{ij}\delta P_{ji}\phi_i(\mathbf{r})\phi_j(\mathbf{r}), \quad (4)$$

since the MO coefficients are fixed.    The equation proved above means that if we obtain the ground-state solution of the effective CI equations (2a-c), the Euler equation for eq. (1a-c) is satisfied.

Therefore, the system of equations given above can be an alternative of the Kohn-Sham DFT.    In particular, the relation given by eq. (3c) is an original part in the iterative CASCI equation.

**Effective one-particle equation for MR-DFT**

In the formulation of MCSCF-DFT, the problem that concerns us now is the variational procedure for molecular orbitals.    The mathematical (not computational) formulation can be archived by a straightforward manner as follows. The first order deviation of the real system is given by

$$\delta E[\rho(\mathbf{r})] = \sum_{ij} \langle \delta i | \hat{H}^{core} + \frac{\delta E_{RC}}{\delta \rho} | j \rangle P_{ji} + \sum_{ijkl} \langle \delta ij | kl \rangle \Pi_{klij} + c.c. \qquad (5)$$

The deviation of the effective MCSCF energy is

$$\begin{aligned}\delta E^{eff} &= \sum_{ij} \langle \delta i | \hat{H}^{core} + \hat{V}_1^{eff} | j \rangle P_{ji} + \sum_{ijkl} \langle \delta ij | kl \rangle \Pi_{klij} + c.c. \\ &+ \sum_{ij} \langle i | \delta \hat{V}_1^{eff} | j \rangle P_{ji} \end{aligned} \qquad (6)$$

which is obviously equivalent to eq. (5) by noting the relations given by eq. (3).    This fact can be confirmed as follows:

$$\begin{aligned}\delta E^{eff} &= \sum_{ij} \langle \delta i | \hat{H}^{core} + \hat{V}_1^{eff} | j \rangle P_{ji} + \sum_{ijkl} \langle \delta ij | kl \rangle \Pi_{klij} + c.c. \\ &+ \sum_{ij} \langle i | \delta \hat{V}_1^{eff} | j \rangle P_{ji} \\ &= \sum_{ij} \langle \delta i | \hat{H}^{core} | j \rangle P_{ji} + \sum_{ijkl} \langle kl | ij \rangle \delta \Pi_{ijkl} + c.c. \\ &+ \int d\mathbf{r} \left[ \delta \rho(\mathbf{r}) V^{RC}(\mathbf{r}) + \rho(\mathbf{r}) \delta V^{RC}(\mathbf{r}) \right] \\ &= \sum_{ij} \langle \delta i | \hat{H}^{core} | j \rangle P_{ji} + \sum_{ijkl} \langle kl | ij \rangle \delta \Pi_{ijkl} + c.c. + \int d\mathbf{r} \frac{\delta E^{RC}(\mathbf{r})}{\delta \rho(\mathbf{r})} \delta \rho(\mathbf{r}) \\ &= \sum_{ij} \langle \delta i | \hat{H}^{core} + \frac{\delta E^{RC}(\mathbf{r})}{\delta \rho(\mathbf{r})} | j \rangle P_{ji} + \sum_{ijkl} \langle kl | ij \rangle \delta \Pi_{ijkl} + c.c. \end{aligned}$$

Note that we here consider the deviations of MOs only, so

$$\delta\rho(\mathbf{r}) = \sum_{ij}\left[P_{ji}\delta\phi_i(\mathbf{r})\phi_j(\mathbf{r}) + c.c.\right].$$

The above equation indicates the equivalence between $\delta E^{eff}$ and $\delta E$ for this density deviation.

The usual treatment of this equation [11] yields the effective one-electron problem:

$$\delta E = \sum_{ij}\langle\delta i|\hat{H}^{core} + \hat{v}^{RC}|j\rangle P_{ji} + \sum_{ijkl}\langle\delta ij|kl\rangle\Pi_{ijkl} + c.c.$$
$$= \sum_{ij}\langle\delta i|F_{ij}|j\rangle + c.c. \quad (7a)$$

$$F_{ij} = \sum_{ij}P_{ji}\left(H^{core}_{ij} + \langle i|\frac{\delta E_{RC}}{\delta\rho}|j\rangle\right) + \sum_{jk,il}\Pi_{jlik}\langle i|V^{pc}|j\rangle \quad (7b)$$

$$\langle i|V^{pc}|j\rangle \equiv \int d\mathbf{r}d\mathbf{r}'\frac{\phi_i^*(\mathbf{r})\phi_l^*(\mathbf{r}')\phi_k(\mathbf{r}')\phi_j(\mathbf{r})}{|\mathbf{r}-\mathbf{r}'|} \quad (7c)$$

Note that the correlation potential for the MCSCF orbitals takes the form similar to that of KS-DFT as

$$v^{RC}(\mathbf{r}) = \frac{\delta E_{RC}}{\delta\rho}(\mathbf{r}). \quad (7d)$$

This is different to that for $V^{RC}(\mathbf{r})$: $V^{RC}(\mathbf{r})$ corresponds to correlation energy per particle while $v^{RC}(\mathbf{r})$ to correlation potential in the context of KS-DFT.

The set of those equation described here is the fundamental equations of MCSCF-DFT.

The algorithm of MCSCF-DFT is as follows:

(i) Set initial MOs and 1-DM.

(ii) Transform integrals to that with the given MOs' basis.

(iii) Compute $V^{RC}(\mathbf{r})$.

(iii) Solve the CASCI-DFT (effective CASCI) equation using $V^{RC}(\mathbf{r})$ to update 1-DM.

(iv) Compute $v^{RC}(\mathbf{r})$.

(v) Solve the CASSCF-DFT (effective one-particle) equation using $v^{RC}(\mathbf{r})$ to update MOs

(vi) If MOs and 1-DM are converged, the computation is completed. Otherwise go back to (ii).

A CASSCF-DFT is a specific version of MCSCF-DFT. The difference between a CASSCF-DFT and a MCSCF is the definition of a residual correlation given by DFT. The computational results are presented elsewhere.